\documentclass[aps,prc,twocolumn,superscriptaddress,amsfonts,amsmath,amssymb,showpacs,floatfix]{revtex4-1}

\usepackage{graphicx}
\usepackage{hyperref}
\hypersetup{colorlinks,citecolor=blue,filecolor=blue,linkcolor=blue,urlcolor=blue}

\begin{document}

\title{Quenching of Single-Particle Strength in {\boldmath$A=15$} Nuclei}

\author{B.~P.~Kay}
\email[E-mail: ]{kay@anl.gov}
\affiliation{Physics Division, Argonne National Laboratory, Lemont, Illinois 60439, USA}
\author{T.~L.~Tang}
\altaffiliation[Present address: ]{Department of Physics, Florida State University, Tallahassee, Florida 32306, USA}
\affiliation{Physics Division, Argonne National Laboratory, Lemont, Illinois 60439, USA}
\author{I.~A.~Tolstukhin}
\affiliation{Physics Division, Argonne National Laboratory, Lemont, Illinois 60439, USA}
\author{G.~B.~Roderick}
\affiliation{Department of Nuclear Physics and Accelerator Applications, Research School of Physics, The Australian National University, Canberra, ACT 2601, Australia}
\author{A.~J.~Mitchell}
\affiliation{Department of Nuclear Physics and Accelerator Applications, Research School of Physics, The Australian National University, Canberra, ACT 2601, Australia}
\author{Y.~Ayyad}
\altaffiliation[Present address: ]{Instituto Galego de F\'{i}sica de Altas Enerx\'{i}as, University of Santiago de Compostela, E-15782 Santiago de Compostela, Spain}
\affiliation{Facility for Rare Isotope Beams, Michigan State University, East Lansing, Michigan 48824, USA}
\author{S.~A.~Bennett}
\affiliation{Schuster Laboratory, The University of Manchester, Oxford Road, Manchester, M13 9PL, UK}
\author{J.~Chen}
\altaffiliation[Present address: ]{Physics Division, Argonne National Laboratory, Lemont, Illinois 60439, USA}
\affiliation{Facility for Rare Isotope Beams, Michigan State University, East Lansing, Michigan 48824, USA}
\author{K.~A.~Chipps}
\affiliation{Physics Division, Oak Ridge National Laboratory, Oak Ridge, Tennessee 37831, USA}
\affiliation{Department of Physics and Astronomy, University of Tennessee, Knoxville, Tennessee 37996, USA}
\author{H.~L.~Crawford}
\affiliation{Nuclear Science Division, Lawrence Berkeley National Laboratory, Berkeley, California 94720, USA}
\author{S.~J.~Freeman}
\affiliation{Schuster Laboratory, The University of Manchester, Oxford Road, Manchester, M13 9PL, UK}
\author{K.~Garrett}
\affiliation{Schuster Laboratory, The University of Manchester, Oxford Road, Manchester, M13 9PL, UK}
\author{M.~D.~Gott}
\affiliation{Physics Division, Argonne National Laboratory, Lemont, Illinois 60439, USA}
\author{M.~R.~Hall}
\affiliation{Physics Division, Oak Ridge National Laboratory, Oak Ridge, Tennessee 37831, USA}
\author{C.~R.~Hoffman}
\affiliation{Physics Division, Argonne National Laboratory, Lemont, Illinois 60439, USA}
\author{H.~Jayatissa}
\affiliation{Physics Division, Argonne National Laboratory, Lemont, Illinois 60439, USA}
\author{A.~O.~Macchiavelli}
\affiliation{Nuclear Science Division, Lawrence Berkeley National Laboratory, Berkeley, California 94720, USA}
\author{P.~T.~MacGregor}
\affiliation{Schuster Laboratory, The University of Manchester, Oxford Road, Manchester, M13 9PL, UK}
\author{D.~K.~Sharp}
\affiliation{Schuster Laboratory, The University of Manchester, Oxford Road, Manchester, M13 9PL, UK}
\author{G.~L.~Wilson}
\affiliation{Department of Physics and Astronomy, Louisiana State University, Baton Rouge, Louisiana 70803, USA}
\affiliation{Physics Division, Argonne National Laboratory, Lemont, Illinois 60439, USA}

\date{\today}

\begin{abstract}

Absolute cross sections for the addition of $s$- and $d$-wave neutrons to $^{14}$C and $^{14}$N have been determined simultaneously via the ($d$,$p$) reaction at 10~MeV/u. The difference between the neutron and proton separation energies, $\Delta S$, is around $-20$~MeV for the $^{14}$C$+$$n$ system and $+8$~MeV for $^{14}$N$+$$n$. The population of the $1s_{1/2}$ and $0d_{5/2}$ orbitals for {\it both} systems is reduced by a factor of approximately 0.5 compared to the independent single-particle model, or about 0.6 when compared to the shell model. This finding strongly contrasts with results deduced from intermediate-energy knockout reactions between similar nuclei on targets of $^{9}$Be and $^{12}$C. The simultaneous technique used removes many systematic uncertainties.

\end{abstract}


\maketitle

{\it Introduction.---}Single-particle motion of nucleons in the nuclear mean field is quenched by correlations, the understanding of which has relevance to the study of nuclear structure, neutron stars, and beyond~\cite{Carlson15,Hen17}. For stable nuclei, the degree of quenching is around 0.6 and appears to be reasonably independent of the reaction mechanism that is used to probe it, the mass of the nucleus, the orbital-angular momentum of the nucleon probed, nucleon separation energy, and is the same within uncertainties for both protons and neutrons~\cite{Kramer01,Kay13,Tostevin21,Aumann21}. 

\begin{figure}[h!]
\centering
\includegraphics[scale=0.7]{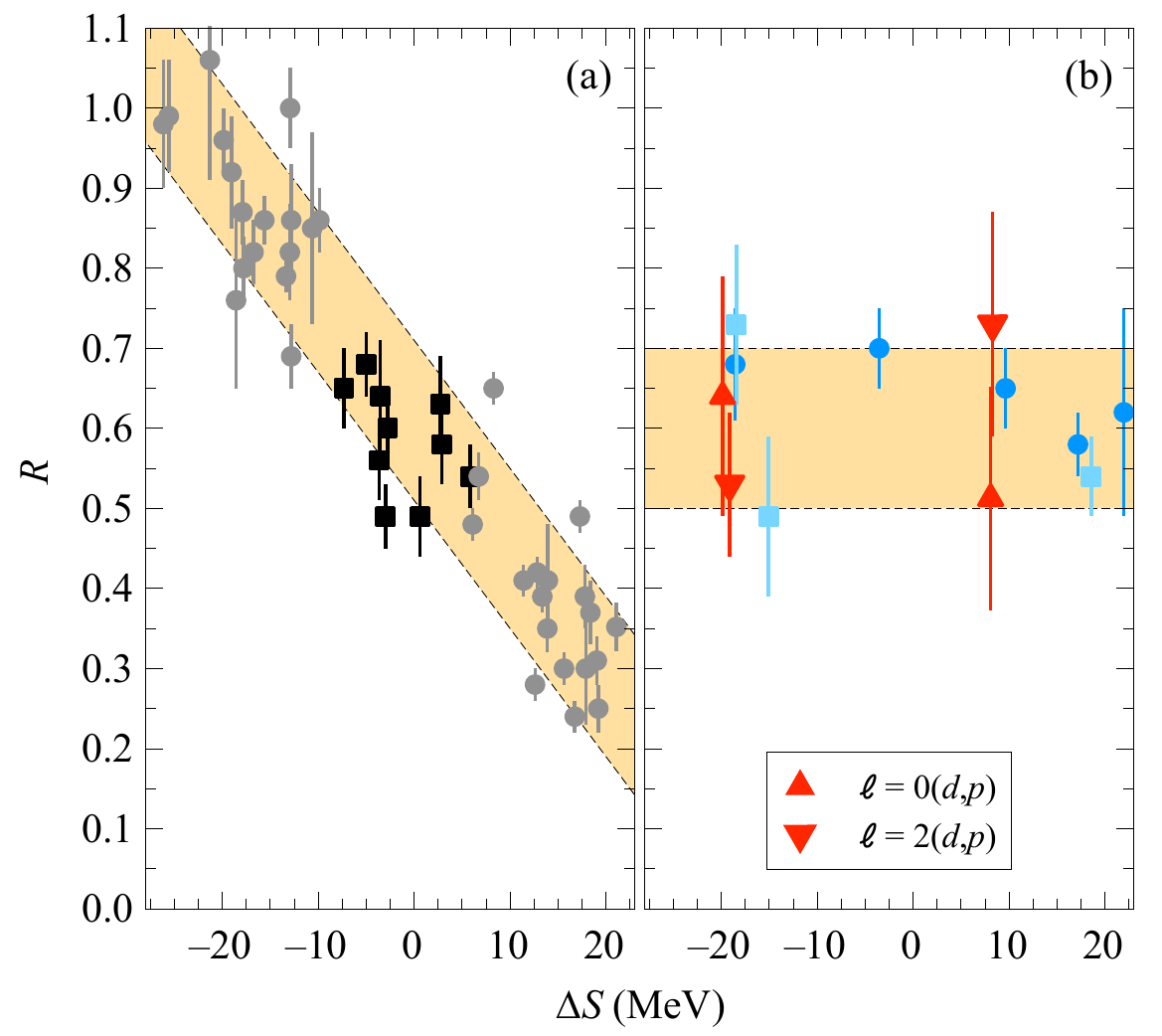}
\caption{\label{fig1} (a) Degree of quenching, $R$, as a function of $\Delta S$ deduced from ($e$,$e'p$) reactions~\cite{Kramer01} (black squares) and from knockout reactions on $^{9}$Be and $^{12}$C targets (grey circles)---data and shaded band from Ref.~\cite{Tostevin21}, compared with (b) results from the current measurement (red triangles) and previous neutron and proton removing transfer reaction study of Ref.~\cite{Flavigny13} (blue squares) and the ($p$,$2p$) study~\cite{Atar18} (blue circles). The shaded band, $R=0.6(1)$, in (b) is to guide the eye. The ($e$,$e'p$) and ($p$,$2p$) measurements are compared to the independent single-particle model and the rest, including the present work, to the shell model.}
\end{figure}

Stable nuclei have differences between their proton ($S_p$) and neutron ($S_n$) separation energies, $\Delta S$, of about $\pm$10~MeV; for proton adding/removing reactions, $\Delta S=S_p^*-S_n$ and for neutrons $S_n^*-S_p$ (the asterisk denotes corrections for excited states where relevant). Information from intermediate-energy heavy-ion knockout reactions on $^{9}$Be and $^{12}$C targets using radioactive-ion beams (referred to as ``HI knockout'' here) has revealed a strong dependence between the degree of quenching (or reduction factor, $R$, defined as the ratio of the experimental and theoretical inclusive cross sections) and $\Delta S$, especially at extreme values of $|\Delta S|>10$~MeV~\cite{Tostevin21}. This trend, shown in Fig.~\ref{fig1}(a), has been the topic of much discussion~\cite{Aumann21}. 

There are limited data on nucleon-removal reactions from such exotic nuclei with different reaction probes, such as transfer and proton knockout, and they appear to disagree with the results of Ref.~\cite{Tostevin21}, for example, the blue points in Fig.~\ref{fig1}(b)~\cite{Flavigny13,Atar18}.

Here, we report on new data, also shown in Fig.~\ref{fig1}(b), from a simultaneous determination of quenching factors for different momentum transfer in the nucleon-adding $^{14}$N($d$,$p$)$^{15}$N and $^{14}$C($d$,$p$)$^{15}$C transfer reactions at energies just above the Coulomb barrier, spanning 28~MeV in $\Delta S$. The ($d$,$p$) reaction is well understood and the simultaneous measurement using these two isobars eliminates many uncertainties. Our result supports the hypothesis that the strong dependence of $R$ on $\Delta S$ seen in the HI-knockout data is unique to that probe.

Some nucleon-removal reaction techniques, such as the ($p$,$2p$) reaction and nucleon-transfer reactions do not yield evidence for this trend in $\Delta S$, but the body of work is very scarce. The neutron-removing $^{34,36,46}$Ar($p$,$d$)$^{33,35,45}$Ar reactions~\cite{Lee10,Manfredi21} (covering a range of $-10\lesssim \Delta S \lesssim +12$~MeV, similar to stable nuclei) and $^{14}$O($d$,$^3$He/$t$)$^{13}$N/$^{13}$O reactions~\cite{Flavigny13} ($\Delta S$ approximately $-18$~MeV and $+18$~MeV, see Fig.~\ref{fig1}(b)) give results that are not necessarily conclusive, but consistent with $R$ being almost constant with  $\Delta S$. More recently, two works using the ($p$,$2p$) knockout reaction on isotopes of oxygen spanning $-20\lesssim \Delta S \lesssim +23$~MeV were published~\cite{Atar18,Kawase18}, also consistent with $R$ being fairly constant with $\Delta S$ (see Fig.~\ref{fig1}(b) for Ref.~\cite{Atar18} results).

From a theoretical perspective, no clear picture has yet emerged concerning the behavior of $R$ as a function of $\Delta S$ in HI knockout, with various efforts made to explore this via nuclear structure and reaction theory (see, for example, Refs.~\cite{Aumann21,Jensen11,Okolowicz16,Paschalis20,Wylie21,Bertulani21,Flavigny12,Lei21}).

To provide additional insights, we took advantage of the weak binding of the $s$ and $d$ states in $^{15}$C, which results in a large, negative value of $\Delta S$. We measured the {\it neutron-adding} ($d$,$p$) reaction at beam energies where the reaction models used are well validated. In early HI-knockout work, Terry {\it et al.}~\cite{Terry04} showed that removal of the $s_{1/2}$ neutron from $^{15}$C, when impinging on a $^9$Be target, had a quenching factor of 0.96(4), in contrast to results for well-bound states near stability.

In order to reduce systematic uncertainties in both the experimental and reaction theory sides, we studied the same reaction simultaneously on two isobars, $^{14}$N($d$,$p$)$^{15}$N and $^{14}$C($d$,$p$)$^{15}$C, under identical conditions, giving an assessment of $R$ at $\Delta S$ of approximately $-$20~MeV and $+$8~MeV (when considering the energy of the excited states populated), and for both $1s_{1/2}$ and $0d_{5/2}$ strength.

\begin{figure}[h]
\centering
\includegraphics[scale=0.47]{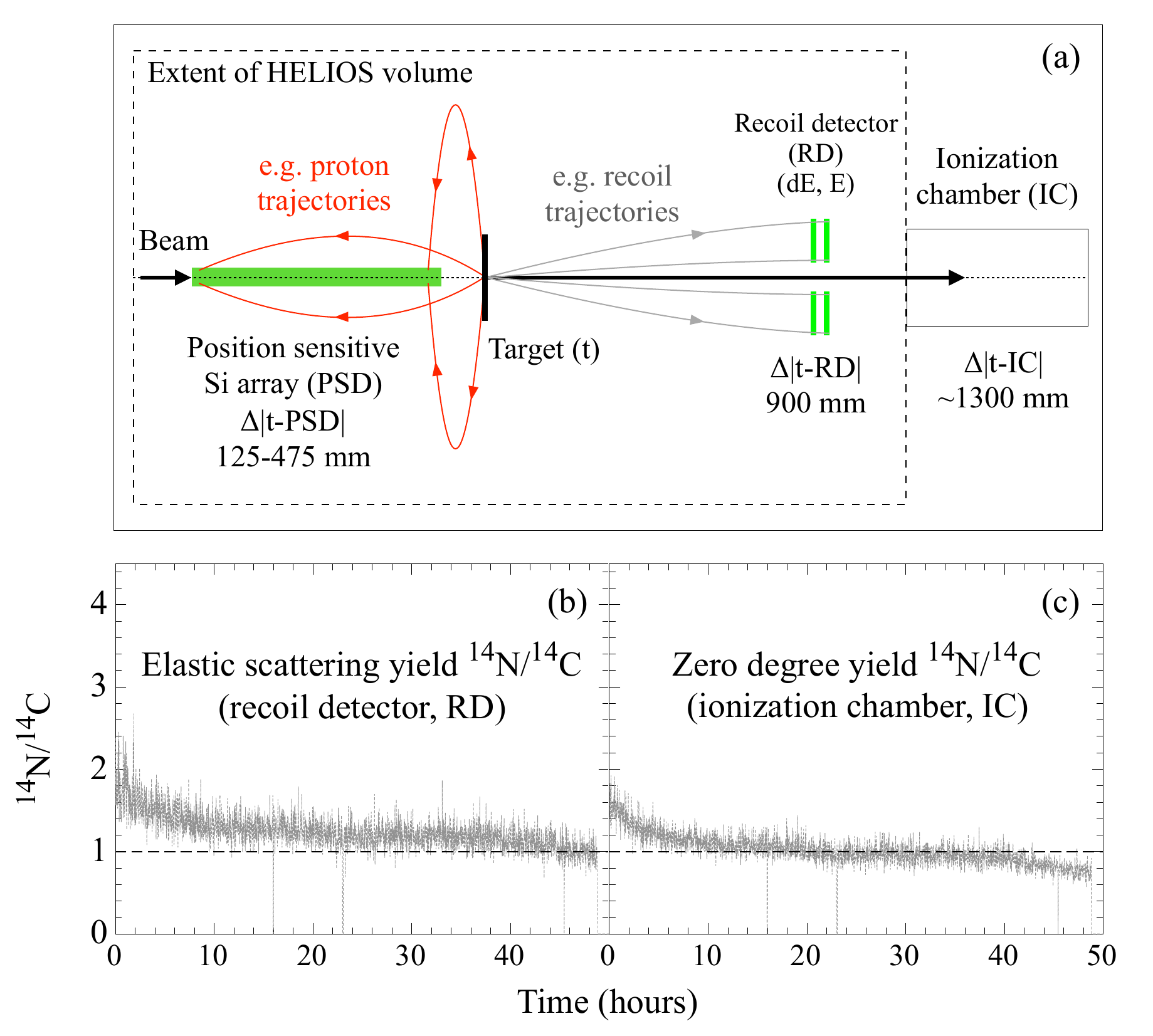}
\caption{\label{fig2} (a) A schematic of the experimental setup and (b) the ratio $^{14}$N/$^{14}$C determined from elastic scattering (raw counts) in the recoil detectors sampled throughout the experiment at a rate of approximately 15~Hz. (c) The same ratio determined from the ionization chamber at the same frequency.}
\end{figure}

Both reactions have been studied before in normal kinematics. Previous results for the $^{14}$C($d$,$p$)$^{15}$C reaction~\cite{Cecil75,Goss75,Murillo94,Mukhamedzhanov11,McCleskey14} are somewhat at odds with each other~\cite{Lee07}, with values of $R$ ranging from $\approx$0.6 to 1.0 for both $1s_{1/2}$ and $0d_{5/2}$ transfer. Several of these measurements suffer from significant contamination present in the $^{14}$C targets and from difficulty in estimating their thickness. Previous measurements of the $^{14}$N($d$,$p$)$^{15}$N reaction~\cite{Philips69,Kretschmer80} are in better agreement with each other; those of Ref.~\cite{Kretschmer80} provide detailed information on the spin and parity using polarized beams. The goal of the current work is to provide an accurate comparisons of the degree of quenching between these two systems lying $\approx$28~MeV apart in $\Delta S$.

{\it Experiment.---}The measurement was carried out in inverse kinematics using the HELIOS spectrometer~\cite{Lighthall10} at the ATLAS facility at Argonne National Laboratory. A cocktail beam of $^{14}$C and $^{14}$N was accelerated to 10~MeV/u at a total intensity ($^{14}$C+$^{14}$N) of approximately 300,000 particles per second. A 124(7)-$\mu$g/cm$^2$ thick deuterated-polyethylene target was used (with an estimated hydrogen content was around 6\%). The thickness was determined offline in five measurements across its surface of $\alpha$-particle energy-loss characteristics, before and after the experiment; the uncertainty is represented by their rms spread. A schematic of the experimental setup, discussed below, is shown in Fig.~\ref{fig2}.

The total beam rate was determined by a fast-counting ionization chamber~\cite{Lai18} located $\approx$1300-mm downstream of the target, where the discriminator rate was used as a scaler to count the total incident ions. The beam composition was sampled at a rate of 15 times per second, triggered (dominantly) by an $\alpha$ source which illuminated the position-sensitive silicon detector array, itself the trigger for the data acquisition. The two different beam species were clearly separated by energy and energy-loss determined measurements in two segments of the ionization chamber. 
\begin{figure}[h]
\centering
\includegraphics[scale=0.7]{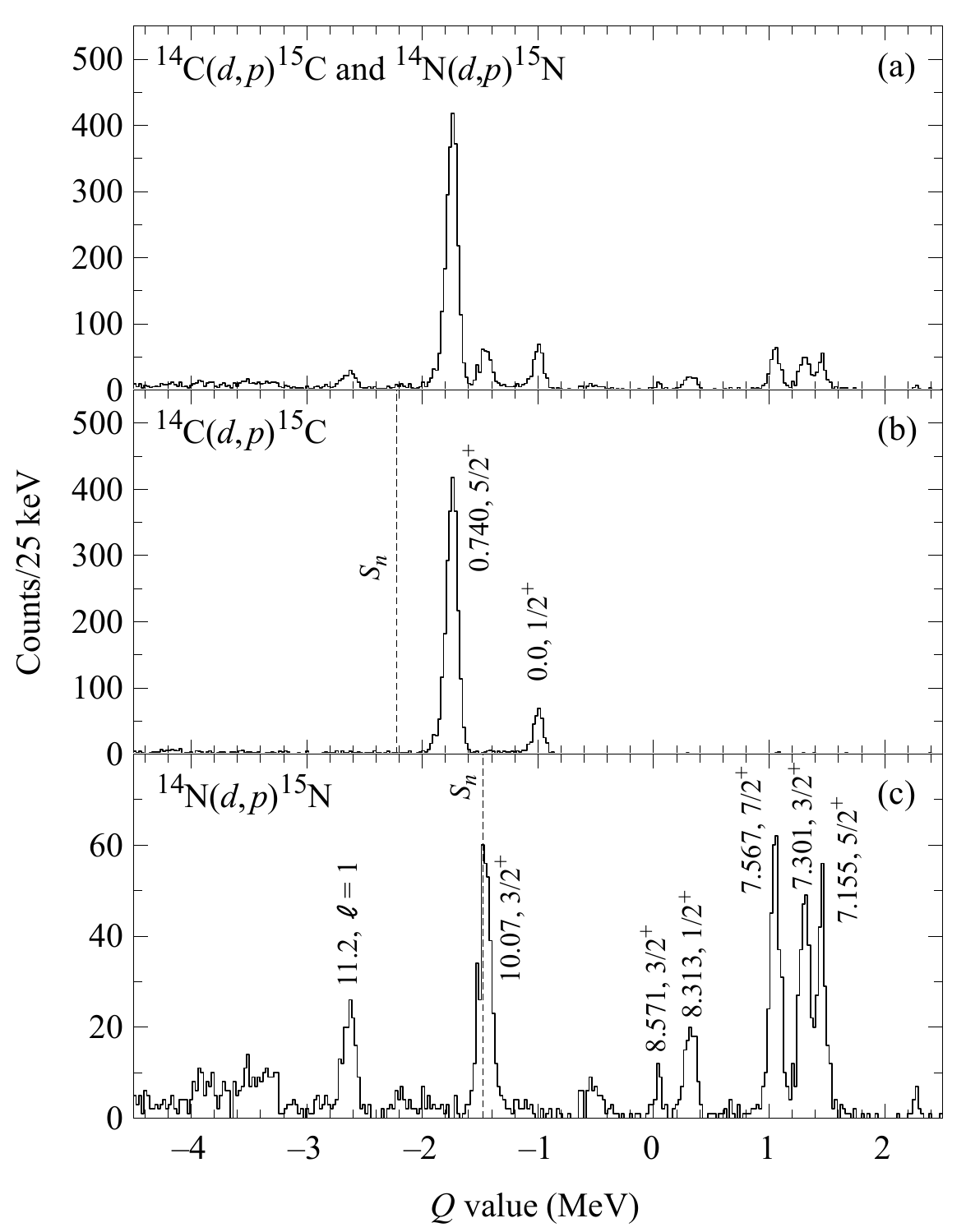}
\caption{\label{fig3} (a) Simultaneously-determined \mbox{$Q$-value} spectrum for the $^{14}$C($d$,$p$)$^{15}$C and $^{14}$N($d$,$p$)$^{15}$N reactions at 10 MeV/u. The recoil-gated reaction yields are shown in Panels (b) and (c). States are labeled by their excitation energy in MeV, and spin or transferred orbital-angular momentum.}
\end{figure}
The beam ratio, determined by this method was corroborated by Rutherford elastic scattering in the recoil detectors, placed 900~mm downstream from the target; a hole in the center allowed unreacted beam to enter the ionization chamber. Both beam-ratio determinations are shown in Fig.~\ref{fig2}, where the ratio determined from the recoil detectors is $\approx$10\% higher in favor of $^{14}$N due to the difference in the elastic-scattering cross sections. From the accumulated yields of $^{14}$N and $^{14}$C, a time-averaged ratio of 1:1 was determined within 3\%, or approximately 1.5$\times$10$^5$~particles per second of both $^{14}$C and $^{14}$N on average.

Protons from the ($d$,$p$) reaction emitted at forward center-of-mass angles (backwards in the laboratory) for both $^{14}$C and $^{14}$N were incident on the Si array, subtending $10^{\circ}\lesssim\theta_{\rm c.m.}\lesssim40^{\circ}$ (the c.m. angle and the detected position along the axis are related to on another~\cite{Wuosmaa07}). The solenoidal $B$ field was 2~T. A $Q$-value resolution of $\approx$125~keV FWHM was achieved, sufficient to separate close-lying states in $^{15}$N. The spectra, shown in Fig.~\ref{fig3}, were recorded simultaneously, and separated by gating on the different recoils. Absolute cross sections were determined from the known target thickness and integrated beam dose in the ionization chamber.


{\it Data reduction and results.---}In $^{15}$C, the $1s_{1/2}$ and $0d_{5/2}$ strength is carried by the ground state and first excited state, respectively, and to first order this represents the $(2j+1)$ vacancy of each. 
\begin{figure}[h]
\centering
\includegraphics[scale=0.98]{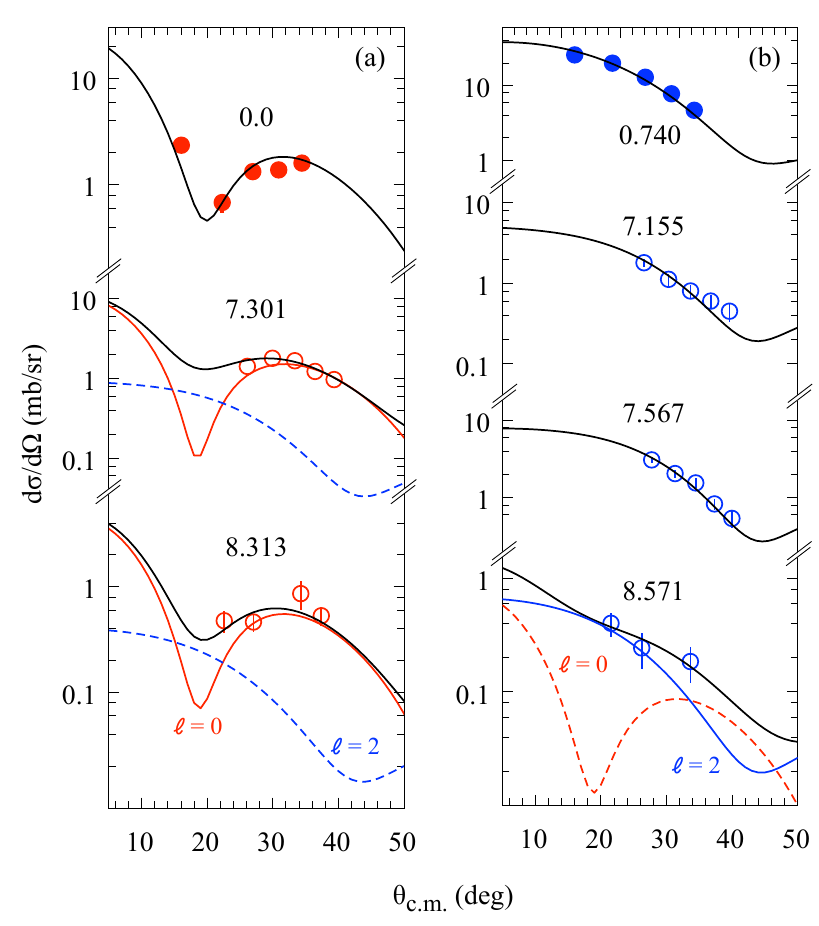}
\caption{\label{fig4} Absolute differential cross sections, labeled by excitation energy in MeV, with fits (solid black lines) determined from a DWBA analysis (see text). The solid data points are for the $^{14}$C($d$,$p$)$^{15}$C reaction and empty symbols for $^{14}$N($d$,$p$)$^{15}$N. Panel (a) shows states populated via (in the case of $^{15}$N, dominantly) $\ell=0$ transfer (red) and (b) for $\ell=2$ (blue). The colored solid (dashed) curves denote the dominant (subdominant) $\ell$-transfer where relevant.}
\end{figure}
Data from the $^{14}$C($p$,$d$)$^{13}$C reaction~\cite{Yasue90} show that about 3\% of the $0d_{5/2}$ strength, and a negligible ($<$1\%) component of the $1s_{1/2}$ strength, is seen in neutron removal (probing occupancy), validating this assumption, and broadly agreeing with shell-model calculations. These results can be readily compared to the inclusive cross sections determined in HI-knockout reactions. 

To extract spectroscopic factors from the experimental data, the distorted-wave Born approximation (DWBA) is used, which has been well validated over many years~\cite{Schiffer12,Kay13}. A similar model, using the adiabatic wave approximation (ADWA) provides a comparison, although at these incident beam energies results are similar. 

The DWBA calculations were done using the finite-range code {\sc Ptolemy}~\cite{ptolemy}, and those for the ADWA calculations used the code {\sc Twofnr}~\cite{twofnr}. In the DWBA case, the deuteron bound state was described by the Argonne $\nu_{18}$ potential~\cite{av18}. For ADWA, the Johnson-Tandy adiabatic model was used~\cite{Johnson74}. The beam species bound-state form factors were generated using a Woods-Saxon potential, defined by $r_0=1.25$~fm, $a=0.65$~fm, $V_{\rm so}=6$~MeV, $r_{\rm so0}=1.1$~fm, and $a_{\rm so}=0.65$~fm. The depth of the potential was varied to reproduce the binding energy of the transferred nucleon to the final state.

Different global optical-model parameterizations were explored in the incoming ($d+^{14}$C/$^{14}$N) and outgoing ($p+^{15}$C/$^{15}$N) channels. For deuterons, these included the parameterizations of Refs.~\cite{An06,Daehnick80,Zhang16}. Those of Ref.~\cite{Daehnick80} were not explicitly derived from fits which included nuclei in this mass range, while those of Refs.~\cite{An06,Zhang16} were. For protons, the parameterizations of Refs.~\cite{Koning03,Varner91} were used. Again, these are not explicitly derived from fits to nuclei in this mass range, but from systems close in mass. For the ADWA calculations, the proton/neutron potentials used are those derived from the nucleon-nucleus optical potentials~\cite{Koning03}. Different combinations of parameter sets and modest variations in the bound-state radius parameter resulted in variations in spectroscopic factors of approximately 15-30\%, which is commensurate with other parameter sensitivity studies~\cite{Flavigny18}. These variations were most significant for the $s$ wave in $^{15}$C and are reflected in the quoted uncertainties. The parameterizations used gave satisfactory fits to previously determined  $^{14}$C($d$,$p$)$^{15}$C data at similar energies~\cite{Goss75,Mukhamedzhanov11,McCleskey14}, further validating the choices made. 

\begin{table}
\caption{\label{tab1} Values of $\Delta S$, DWBA ($SF$) and shell-model ($SF_{\rm SM}$) spectroscopic factors, and $R$ for the $1s_{1/2}$ and $0d_{5/2}$ strength in $^{15}$C and $^{15}$N.}
\newcommand\T{\rule{0pt}{3ex}}
\newcommand \B{\rule[-1.2ex]{0pt}{0pt}}
\begin{ruledtabular}
\begin{tabular}{lccccc}
\T\B$^AX$ \B & $nlj$ & $\Delta S$ (MeV) & $SF$ & $SF_{\rm SM}$ & $R$   \\
\hline
$^{15}$C\T	 &	$1s_{1/2}$	&	$-$19.86	& 0.51(12) 	& 0.80 & 0.64(15) 	\\
\T	&		$0d_{5/2}$	&	$-$19.12	& 0.41(7)	& 0.78 & 0.53(9)   \\
$^{15}$N\T	&	$1s_{1/2}$	&	$+$8.08	& 0.41(11) & 0.80 & 0.51(14)	 	\\
\T     &		$0d_{5/2}$	&	$+$8.29	& 0.61(12) & 0.84 & 0.73(14)	\\
\end{tabular}
\end{ruledtabular}
\end{table}

The spectroscopic factors were derived from fitting the experimental angular distributions, illustrated in Fig.~\ref{fig4}. The deduced $R$ values are given in Table~\ref{tab1}, derived by comparing the DWBA-derived spectroscopic factors with shell-model calculations carried out using {\sc oxbash}~\cite{BrownOx} and the WBP interaction~\cite{Warburton92}.

For $^{15}$C, $R$ is determined to be 0.64(15) and 0.53(9) for the $1s_{1/2}$ and $0d_{5/2}$ strength, respectively. The uncertainties are estimated by the rms spread in the different DWBA and ADWA fits. This contrasts the value of 0.96(4) for the $s$-wave strength deduced in the HI-knockout study~\cite{Terry04}.

For $^{15}$N, the $1s_{1/2}$ and $0d_{5/2}$ strength is carried by two sets of multiplets, which were previously studied~\cite{Philips69,Kretschmer80}. The majority of the strength ($\approx$95\%) is shared among five states between 7.155 and 8.571~MeV in excitation energy, naively arising from the coupling of a $1s_{1/2}$ or $0d_{5/2}$ neutron to the 1$^+$, $T=0$ $^{14}$N ground state. A lower-lying, positive-parity doublet (5.270 and 5.299~MeV) carries the remaining $<$5\% of the strength, with small spectroscopic factors seen in previous transfer-reaction studies~\cite{Kretschmer80}. In this measurement, they were not observed due to the detector acceptance. The omission of these states from the summed strength contributes to the overall uncertainty to a similar extent as other sources. There is no other $1s_{1/2}$ or $0d_{5/2}$ strength below the neutron-separation threshold. As for $^{14}$C, the $1s_{1/2}$ and $0d_{5/2}$ orbitals in $^{14}$N are approximately fully unoccupied, with no strength observed in neutron-removal reactions~\cite{Hinterberger67}, corroborated by the shell-model calculations.

The $^{15}$N $0d_{5/2}$ strength is shared between two pure $0d_{5/2}$ states and one mixed state, as was well established in previous studies (e.g., Ref.~\cite{Kretschmer80}). The \mbox{$J^{\pi}=3/2^+$} state at 8.571~MeV has strength that is approximately one-third $1s_{1/2}$ and two-thirds $0d_{5/2}$ in terms of $S$ and, though weakly populated, represents about 10\% of the total $0d_{5/2}$ strength. The remaining strength is carried by the $J^{\pi}=5/2^+$ and $7/2^+$ states at 7.155 and 7.567~MeV. Most of the $1s_{1/2}$ strength lies in the 7.301~MeV, $J^{\pi}=3/2^+$ state (which has small admixtures of $0d_{5/2}$ strength) and the 8.313-MeV, $J^{\pi}=1/2^+$ state (mixed with a small fraction of $\ell=2$, $0d_{3/2}$ strength as determined from a previous ($d$,$p$) study with polarized beams~\cite{Kretschmer80}). The remaining $1s_{1/2}$ strength is carried by the 8.571-MeV state mentioned above. The $SF$ values for $^{15}$N (see Table~\ref{tab1} and in Fig.~\ref{fig1}) represent the summed spectroscopic factors for the $1s_{1/2}$ and $0d_{5/2}$ orbitals. We note that the states discussed here account for all the $1s_{1/2}$ and $0d_{5/2}$ strength below the neutron-separation energy threshold and, as for $^{15}$C, can be readily compared to the inclusive cross sections determined in HI-knockout reactions. The larger uncertainties for $^{15}$N are a consequence of fitting the $\ell=0+2$ states.


{\it Conclusions.---}In a measurement where many systematic errors are absent because of the ratio method, we show that for two different nuclei separated by $\approx$28~MeV in $\Delta S$, a similar degree of quenching of the single-particle strength is observed for both the $s$ and $d$ excitations. This contrasts with the factor-of-two difference observed in HI-knockout over the same range in $\Delta S$. This work provides the first assessment of {\it neutron-adding} ($d$,$p$)-reaction data at extreme $\Delta S$, complementing a very limited number of data sets from nucleon-removing knockout and transfer reactions, adding to the growing evidence of a discrepancy between HI knockout and results using other probes.

There is an expectation that more data will become available as radioactive beams at energies above the Coulomb barrier become more widely available for transfer-reaction studies, for example, at the Facility for Rare Isotope Beams. With many of these facilities likely to provide isobarically {\it impure} beams, the simultaneous beams technique discussed in this work could be a convenient and useful approach for future studies. Developing a body of transfer-reaction data, using both adding and removing reactions, comparable in magnitude to that of the HI-knockout data is likely essential to guide theoretical insights into this discrepancy.

{\it Acknowledgements.---}The authors recognize fruitful discussions with John Schiffer and would like to acknowledge Bob Scott and the ATLAS operations team for the delivery of the cocktail beam. This material is based upon work supported by the U.S. Department of Energy, Office of Science, Office of Nuclear Physics, under Contracts No. DE-AC02-06CH11357 (Argonne), DE-AC02-05CH11231 (Berkeley), DE-AC05-00OR22725 (Oak Ridge), and DE-FG02-96ER40978 (Louisiana), the International Technology Center Pacific (ITC-PAC) under Contract No. FA520919PA138, the UK Science and Technology Facilities Council Grant Nos. ST/P004423/1 and No. ST/T004797/1, and the U.S. National Science Foundation (NSF) under Cooperative Agreement No. PHY-1565546. This research used resources of ANL’s ATLAS facility, which is a DOE Office of Science User Facility.


\end{document}